# Panel flutter characteristics of sandwich plates with CNT reinforced facesheets using an accurate higher-order theory

A. Sankar[a,c,2], S. Natarajan[b,1], M. Haboussi[d], K. Ramajeyathilagam[c], M. Ganapathi[a]

[a]Tech Mahindra Ltd., Electronic City, Bangalore- 560 100, India.
[b]School of Civil and Environmental Engineering, The University of New South Wales, Sydney, NSW 2052, Australia.
[c]School of Aeronautical Sciences, Hindustan University, Keelambakkam, Chennai- 603103, India.
[d]Université Paris 13-CNRS, LSPM, UPR 3407, Villetaneuse, F-93430, France.

**Abstract**

In this paper, the flutter characteristics of sandwich panels with carbon nanotube (CNT) reinforced face sheets are investigated using QUAD-8 shear flexible element developed based on higher-order structural theory. The formulation accounts for the realistic variation of the displacements through the thickness, the possible discontinuity in the slope at the interface, and the thickness stretch affecting the transverse deflection. The in-plane and rotary inertia terms are also included in the formulation. The first-order high Mach number approximation to linear potential flow theory is employed for evaluating the aerodynamic pressure. The solutions of the complex eigenvalue problem, developed based on Lagrange's equation of motion are obtained using the standard method for finding the eigenvalues. The accuracy of the present formulation is demonstrated considering the problems for which solutions are available. A detailed numerical study is carried out to bring out the efficacy of the higher-order model over the first-order theory and also to examine the influence of the volume fraction of the CNT, core-to-face sheet thickness, the plate thickness and the aspect ratio, damping and the temperature on the flutter boundaries and the associated vibration modes.

[1]School of Civil and Environmental Engineering, The University of New South Wales, Sydney, NSW 2052, Australia.
Tel: +61293855030; E-mail: s.natarajan@unsw.edu.au; snatarajan@cardiffalumni.org.uk
[2]Research Scholar

## 1. Introduction

In recent years, non-structured, non-metallic materials have spurred considerable interest in the materials community partly because of their potential for large gains in mechanical and physical properties as compared to standard structural materials. In particular, carbon nanotube/polymer composites may provide order-of-magnitude increase in the strength and the stiffness when compared to typical carbon fiber/polymer composites [1]. Due to these reasons, structures made of such materials have great potentials in the construction of future supersonic /hypersonic space vehicles and reusable transportation systems. Among the various structural constructions, the sandwich type of structures are more attractive due to their outstanding bending rigidity, low specific weight, excellent vibration characteristics and good fatigue properties. These sandwich constructions can be a candidature for the requirement of lightweight and high bending stiffness in the design. A typical sandwich structure may consist of a homogeneous core with facesheets. To improve the characteristics of these structures, the facesheets can be laminated composites [2], functionally graded materials [3] or polymer matrix with reinforcements [4]. The definite advantages offered by the carbon nanotube reinforced composites (CNTRCs) over the carbon fibre-reinforced composites have prompted the engineers to design and analyse sandwich structures with CNTRC facings [5].

Some studies conducted in evaluating the mechanical properties of CNTs can be seen in the literature [6, 7]. Thostenson and Chou [6] showed that the addition of nanotubes increases the tensile modulus, the yield strength and the ultimate strength of the polymer films. Their study has also brought out that the polymer films with aligned nanotubes as reinforcements yield superior strength when compared to randomly oriented nanotubes. The properties of the polymer films can also be optimized by varying the distribution of CNTs through the thickness of the film. Formica *et al.,* [7] highlighted that the CNT reinforced plates can be tailored to respond to an external excitation. These experimental investigations have created great interest among structural modeling and simulation analysts. For predicting the realistic behavior of sandwich structures with CNTRC facings, more accurate analytical/numerical models based on the three-dimensional models may be computationally involved and expensive. Hence, among the researchers, there is a growing appreciation of the importance of applying two-dimensional theories with new kinematics for the evolution of the accurate structural analysis. Few important contributions pertaining to the sandwich plates with CNTRC facesheets and the structural theories proposed for the analysis of such structures are discussed here. Based on the first-order shear deformation theory, Zhu *et al.,* [8] studied the static and free vibration of CNT reinforced plates. They considered polymer matrix with CNT reinforcement, neglecting the temperature effects. It was predicted that the CNT volume fraction has greater influence on the fundamental frequency and the maximum center deflection. Wang and Shen [9] studied the large amplitude vibration of nano-composite plates resting on the elastic foundation using a perturbation technique. The governing equations were based on simple higher-order shear deformation theory. Their study brought out that while the linear frequencies decrease with the addition of CNTs, the nonlinear to linear frequency ratio increased, especially when increasing the temperature or by decreasing the foundation stiffness. Arani *et al.,* [10], Liew *et al.,* [11] and Lei *et al.,* [12] studied the buckling and post-buckling characteristics of CNT reinforced plates using the finite element and meshless methods, respectively. It was revealed that the reinforcement with CNT increases the load carrying capacity of the plate. Aragh *et al.,* [13] used the generalized differential quadrature method and obtained a semi-analytical solution for 3D vibration of cylindrical panels. It was shown that graded CNTs with symmetric distribution through the thickness have high capabilities to alter the natural frequencies when compared to the uniformly distributed or asymmetrically distributed.

It is observed from these investigations that first- order shear deformation theory has been widely employed for the static and free vibration analyses of CNT reinforced plates by many researchers whereas the simplified higher-order model considering variation in in-plane displacements has been used by few authors. However, the available literature pertaining to sandwich structures with CNT reinforced facesheets is rather limited compared to those of fibre-reinforced facings plates. Various theories and structural models such as global-local finite element model using hierarchical multiple assumed displacement fields [14], generalized multiscale plate theories [15], variational asymptotic structural models [16], generalized unified formulation with zig-zag theory [17], etc. that account for the variation of in-plane/transverse displacement through the thickness have been employed for investigating the structural behavior of laminated reinforced composite structures. In this context, Ali *et al.,* [18] and Ganapathi and Makhecha [19] have used a higher-order plate theory based on global approach for multi-layered laminated composites by incorporating the realistic through the thickness approximations of the in-plane and transverse displacements by adding a zig-zag function and higher-order terms, respectively. This approach has proved to give very accurate results and computationally less expensive for the composite laminates compared to those of layerwise theory in which the number of unknowns to be solved increases with the increase in the number of mathematical or physical layers. Such higher-order model for the study of sandwich plates with CNT reinforced facesheets may be worthwhile to consider as a candidature while comparing with the other formulations available in the literature.

The increased effort towards integrating these materials in the construction of aerospace structures has necessitated investigating the aeroelastic stability issues of such structures. The panel flutter phenomenon is one of the aeroelastic dynamic instability problems encountered in the flight of aerospace vehicles. It is the self-excited oscillation of the external skin of a flight vehicle when exposed to airflow along its surface. A comprehensive review of the theory associated with panel flutter analysis can be had from several articles such as Refs. [20-22]. This study pertaining to composite laminates and functionally graded material structures constituting metal/and ceramic has received considerable attention in the literature [23-25]. However, this type of analysis is not accomplished in the literature considering sandwich panels with CNTRC facings and it is worth investigating flutter stability characteristics of such structures exposed to aerodynamic flow.

*Approach.*

In this paper, a $C^0$ 8-noded quadrilateral plate element with 13 degrees of freedom per node [19, 26, 27] based on the higher order theory [18] is employed to study the flutter analysis of thick/thin sandwich plates with carbon nanotube reinforced facesheets. The aerodynamic force is evaluated assuming the first-order High Mach number approximation to linear potential theory. The efficacy of the present formulation is illustrated through the numerical studies by various structural models deduced from the present higher-order theory considering parameters such as CNT volume fraction, core-to-facesheets thickness ratio, plate thickness and aspect ratios, and temperature. The influence of coalescence modes determining the flutter boundary is also discussed.

*Outline.*

The paper is organized as follows. The computation of the effective properties of carbon nanotube reinforced composites is discussed in the next section. Section 3 presents the higher order accurate theory to describe the plate kinematics and Section 4 describes the 8-noded quadrilateral plate element employed in this study. The numerical results for the aeroelastic stability of thick/thin sandwich carbon nanotube reinforced functionally graded plates are given in Section 5, followed by concluding remarks in the last section.

**2. Theoretical Formulation**

Consider a CNT reinforced sandwich plate with the coordinate system *x, y, z* which has its origin at the corner of the plate on the middle plane as shown in Figure 1. The length, the width and the total thickness of the plate are *a, b* and *h*. The thickness of each CNT reinforced facesheet is $h_f$ and the thickness of homogeneous core layer is $h_H$. It is assumed that the CNT reinforced layer is made from a mixture of single walled CNT with uniformly distributed or functionally graded in the thickness direction and the matrix is assumed to be isotropic. The effective properties of such reinforced structures can be computed by Mori-Tanaka scheme [13] or by the rule of mixtures. As the rule of mixture is simple, it is employed here to estimate the overall material properties of the structures. According to extended rule of mixtures, the effective material properties of the CNT reinforced matrix are given by [28]:

$$E_{11} = \eta_1 V_{CN} E_{11}^{CN} + V_m E_m$$

$$\frac{\eta_2}{E_{22}} = \frac{V_{CN}}{E_{22}^{CN}} + \frac{V_m}{E_m}$$

$$\frac{\eta_3}{G_{12}} = \frac{V_{CN}}{G_{12}^{CN}} + \frac{V_m}{G_m}$$

$$\nu_{12}^{CN} = \nu_{12}^{CN} V_{CN}^* + \nu_m V_m$$

$$\rho = \rho_{CN} V_{CN} + \rho_m V_m \tag{1}$$

where, $E_{11}^{CN}$, $E_{22}^{CN}$ and $G_{12}^{CN}$ are the Young's moduli and the shear modulus of CNT, respectively. $E_m$ and $G_m$ are corresponding properties of the matrix. The CNT efficiency parameters ($\eta_1, \eta_2, \eta_3$) are introduced to account for the inconsistency in the load transfer between the CNT and the matrix. The values of the efficiency parameters are obtained by matching the elastic modulus of the CNT reinforced polymer matrix from the MD stimulation results with the numerical results obtained from the rule of mixtures. $V_{CN}$ and $V_m$ are the volume fraction of the CNT and the matrix, respectively and they are related by $V_{CN} + V_m = 1$.

The CNT distributions in the facesheet are functionally graded by linearly varying the volume fraction of the CNT in the thickness direction. It is assumed the volume fraction $V_{CN}$ for the top face sheet as

$$V_{CN} = 2\left(\frac{t_1 - Z}{t_1 - t_0}\right)V_{CN}^*$$

and for the bottom facesheet as

$$V_{CN} = 2\left(\frac{Z - t_2}{t_3 - t_2}\right)V_{CN}^* \tag{2}$$

where,

$$V_{CN}^* = \frac{w_{CN}}{w_{CN} + \left(\frac{\rho_{CN}}{\rho_m}\right)[1 - w_{CN}]} \tag{3}$$

where $w_{CN}$ is the mass fraction of the nanotube, $\rho_{CN}$ and $\rho_m$ are the mass densities of the CNT and the matrix, respectively. The thermal expansion coefficient in the longitudinal and the transverse directions can be expressed as [28]:

$$\alpha_{11} = \alpha_{11}^{CN} V_{CN} + \alpha_m V_m$$
$$\alpha_{22} = (1 + \nu_{12}^{CN})\alpha_{22}^{CN} V_{CN} + (1 + \nu_m)\alpha_m V_m - \nu_{12}\alpha_{11} \tag{4}$$

where, $\alpha_{11}^{CN}$, $\alpha_{22}^{CN}$ and $\alpha_m$ are the thermal expansion coefficients for the CNT and the matrix, respectively and $\nu_{12}^{CN}$ and $\nu_m$ are the Poisson's ratio.

## 3. Governing differential equations

The sandwich plate is assumed to be made of three discrete layers with a homogeneous core. The in-plane displacements $u^k$ and $v^k$, and the transverse displacement $w^k$ for the $k^{th}$ layer, are assumed as [18, 26]:

$$u^k(x,y,z,t) = u_o(x,y,t) + z\theta_x(x,y,t) + z^2\beta_x(x,y,t) + z^3\phi_x(x,y,t) + S^k\psi_x(x,y,t)$$

$$v^k(x,y,z,t) = v_o(x,y,t) + z\theta_y(x,y,t) + z^2\beta_y(x,y,t) + z^3\phi_y(x,y,t) + S^k\psi_y(x,y,t)$$

$$w^k(x,y,z,t) = w_o(x,y,t) + zw_1(x,y,t) + z^2\Gamma(x,y,t) \tag{5}$$

The terms with even powers in $z$ in the in-plane displacements and the odd powers of $z$ occurring in the expansion for $w^k$ correspond to the stretching problem. However, the terms with odd powers of $z$ in the in-plane displacements and the even ones in the expression for $w^k$ represent the flexure problem. $u_0, v_0, w_0$ are the displacements of a generic point on the reference surface; $\theta_x, \theta_y$ are the rotations of normal to the reference surface about the $y$ and $x$ axes, respectively; $w_1, \beta_x, \beta_y, \Gamma, \phi_x, \phi_y$ are the higher order terms in the Taylor's series expansions, defined at the reference surface. $\psi_x$ and $\psi_y$ are generalized variables associated with the zigzag function, $S^k$. The zigzag function, $S^k$ as given in [29, 19, 17] is defined by

$$S^k = 2(-1)^k \frac{z_k}{h_k} \tag{6}$$

where $z_k$ is the local transverse coordinate with the origin at the center of the $k^{th}$ layer and $h_k$ is the corresponding layer thickness. Thus, the zigzag function is piecewise linear with values of $-1$ and $1$ alternatively at different interfaces. The 'zigzag' function, as defined above, takes care of the inclusion of the slope discontinuities of $u$ and $v$ at the interfaces of the sandwich plate as observed in the exact three-dimensional elasticity solutions of thick laminates. The main advantage of using such a function in the formulation is more economical than a discrete layer approach [30, 31]. Although both these approaches account for the slope discontinuity at the interfaces, the number of unknowns increases with the increase in the number of layers in the discrete layer approach, whereas it remains constant in the present approach.

The strains in terms of mid-plane deformation, rotations of normal, and higher order terms associated with displacements are as,

$$\{\varepsilon\} = \begin{Bmatrix} \varepsilon_{bm} \\ \varepsilon_s \end{Bmatrix} \tag{7}$$

The vector $\{\varepsilon_{bm}\}$ includes the bending and the membrane terms of the strain components and the vector $\{\varepsilon_s\}$ contains the transverse shear strain terms. These strain vectors can be defined as

$$\{\varepsilon_{bm}\} = \begin{Bmatrix} \varepsilon_{xx} \\ \varepsilon_{yy} \\ \varepsilon_{zz} \\ \gamma_{xy} \end{Bmatrix} = \begin{Bmatrix} u_{,x} \\ v_{,y} \\ w_{,x} \\ u_{,y} + v_{,x} \end{Bmatrix} = \varepsilon_0 + z\varepsilon_1 + z^2 \varepsilon_2 + z^3 \varepsilon_3 + S^k \varepsilon_4 \tag{8}$$

$$\{\varepsilon_s\} = \begin{Bmatrix} \gamma_{xz} \\ \gamma_{yz} \end{Bmatrix} = \begin{Bmatrix} u_{,z} + w_{,x} \\ v_{,z} + w_{,y} \end{Bmatrix} = \gamma_0 + z\gamma_1 + z^2 \gamma_2 + S^k_{,z} \gamma_3 \tag{9}$$

where,

$$\{\varepsilon_0\} = \begin{Bmatrix} u_{0,x} \\ v_{0,y} \\ w_1 \\ u_{0,y} + v_{0,x} \end{Bmatrix}, \{\varepsilon_1\} = \begin{Bmatrix} \theta_{x,x} \\ \theta_{y,y} \\ 2\Gamma \\ \theta_{x,y} + \theta_{y,x} \end{Bmatrix}, \{\varepsilon_2\} = \begin{Bmatrix} \beta_{x,x} \\ \beta_{y,y} \\ 0 \\ \beta_{x,y} + \beta_{y,x} \end{Bmatrix}, \{\varepsilon_3\} = \begin{Bmatrix} \phi_{x,x} \\ \phi_{y,y} \\ 0 \\ \phi_{x,y} + \phi_{y,x} \end{Bmatrix},$$

$$\{\varepsilon_4\} = \begin{Bmatrix} \psi_{x,x} \\ \psi_{y,y} \\ 0 \\ \psi_{x,y} + \psi_{y,x} \end{Bmatrix} \tag{10}$$

and

$$\{\gamma_0\} = \begin{Bmatrix} \theta_x + w_{0,x} \\ \theta_y + w_{0,y} \end{Bmatrix}, \{\gamma_1\} = \begin{Bmatrix} 2\beta_x + w_{1,x} \\ 2\beta_y + w_{1,y} \end{Bmatrix}, \{\gamma_2\} = \begin{Bmatrix} 3\phi_x + \Gamma_{,x} \\ 3\phi_y + \Gamma_{,y} \end{Bmatrix}, \{\gamma_3\} = \begin{Bmatrix} \psi_x S^k_{,z} \\ \psi_y S^k_{,z} \end{Bmatrix} \quad (11)$$

The subscript comma denotes partial derivatives with respect to the spatial coordinate succeeding it. The constitutive relations for an arbitary layer $k$ can be expressed as:

$$\begin{aligned} \sigma &= \{\sigma_{xx} \quad \sigma_{yy} \quad \sigma_{xy} \quad \sigma_{zz} \quad \sigma_{xz} \quad \sigma_{yz}\}^T \\ &= Q^k \{\varepsilon_{bm} \quad \varepsilon_s\}^T \end{aligned} \quad (12)$$

where $Q_k$ is the stiffness matrix defined as

$$Q^k_{11} = \frac{E_{11}}{1 - \upsilon_{12}\upsilon_{21}}; \quad Q^k_{22} = \frac{E_{22}}{1 - \upsilon_{12}\upsilon_{21}}; \quad Q^k_{12} = \frac{\upsilon_{21} E_{11}}{1 - \upsilon_{12}\upsilon_{21}};$$

$$Q^k_{44} = G_{23}; \quad Q^k_{55} = G_{13}; \quad Q^k_{66} = G_{12}; \quad Q^k_{16} = Q^k_{26} = 0 \quad (13)$$

For the homogeneous core, the shear modulus G is related to the Young's modulus by: $E = 2G(1+\upsilon)$.

The governing equations are obtained by applying the Lagrangian equations of motion given by

$$\frac{d}{dt}\left[\frac{\partial(T-U)}{\partial \dot{\delta}_i}\right] - \left[\frac{\partial(T-U)}{\partial \delta_i}\right] = 0, \quad i = 1, 2 \ldots n \quad (14)$$

where $\delta_i$ the vector of degrees of freedom and $T$ is the kinetic energy of the sandwich plate given by;

$$T(\delta) = \frac{1}{2} \iint \left[\sum_{k=1}^n \int_{h_k}^{h_{k+1}} \rho_k \{\dot{u}_k \quad \dot{v}_k \quad \dot{w}_k\}\{\dot{u}_k \quad \dot{v}_k \quad \dot{w}_k\}^T dz\right] dxdy \quad (15)$$

where $\rho_k$ is the mass density of the $k^{th}$ layer, $h_k$, and $h_{k+1}$ are the $z$ coordinates to the bottom and top surfaces of the $k$th layer. The potential energy functional $U$ is given by,

$$U(\delta) = \frac{1}{2} \iint \left[\sum_{k=1}^n \int_{h_k}^{h_{k+1}} \sigma^T \varepsilon \, dz\right] dxdy - W_a(\delta) \quad (16)$$

The work done by the applied non-conservative load is

$$W_a(\delta) = \iint \Delta p \, w \, dx \, dy \quad (17)$$

where $\Delta p$ is the aerodynamic pressure. The aerodynamic pressure based on first-order high Mach number approximation to linear potential flow is given as [20-22]

$$\Delta p = \frac{\rho_a U_a^2}{\sqrt{M_\infty^2 - 1}} \left[\frac{\partial w}{\partial x} + \frac{1}{U_a}\left(\frac{M_\infty^2 - 2}{M_\infty^2 - 1}\right)\frac{\partial w}{\partial t}\right] \quad (18)$$

where $\rho_a$, $U_a$, and $M_\infty$ are the free stream air density, velocity of air and the Mach number, respectively. Substituting equations (15) to (18) in Lagrange's equations of motion, the following governing equation is obtained:

$$M\ddot{\delta} + g_T D_A \dot{\delta} + (K + \lambda^* \bar{A})\delta = 0 \tag{19}$$

where $K$ is the stiffness matrix, $M$ is the consistent mass matrix, $\lambda^* = \dfrac{\rho_a U_a^2}{\sqrt{M_\infty^2 - 1}}$, $\bar{A}$ is the aerodynamic force matrix and $g_T = \dfrac{\lambda^*(M_\infty^2 - 1)}{U_a(M_\infty^2 - 1)}$ is the aerodynamic damping parameter, the damping matrix $D_A$ can be considered as the scalar multiple of mass matrix by neglecting the shear and rotary inertia terms of the mass matrix $M$ and after substituting the characteristic of the time function $\ddot{\delta} = -\omega^2 \delta$, the following algebraic equation is obtained:

$$\left[\left([K] + \lambda^*[\bar{A}]\right) - \bar{k}[M]\right]\delta = 0 \tag{20}$$

where the eigenvalue $\bar{k} = -\omega^2 - (g_T \omega/\rho h)$ includes the contribution of the aerodynamic damping. Equation (20) is solved for eigenvalues for a given value of $\lambda^*$. In the absence of aerodynamic damping, i.e., when $\lambda^* = 0$, the eigenvalue of $\omega$ is real and positive, since the stiffness matrix and the mass matrix are symmetric and positive definite. However, the aerodynamic matrix $\bar{A}$ is unsymmetric and hence complex eigenvalues $\omega$ are expected when $\lambda^* > 0$. As $\lambda^*$ increases monotonically from zero, two of these eigenvalues will approach each other and become complex conjugates. In this study, $\lambda^*_{cr}$ is considered to be the value of $\lambda^*$ at which the first coalescence occurs. In the presence of aerodynamic damping, the eigenvalues $\bar{k}$, in equation (20) becomes complex with increase in the value of $\lambda^*$. The corresponding frequency can be written as:

$$\bar{k} = -\omega^2 - (g_T \omega/\rho h) = \bar{k}_R - i\bar{k}_I \tag{21}$$

where the subscripts $R$ and $I$ refer to the real and the imaginary part of the eigenvalue. The flutter boundary is reached ($\lambda^* = \lambda^*_{cr}$), when the frequency $\omega$ becomes pure imaginary number, i.e. $\omega = i\sqrt{\bar{k}_R}$ at $g_T = \bar{k}_I/\sqrt{\bar{k}_R}$. In practice, the value of $\lambda^*_{cr}$ is determined from a plot of $\omega_R$ vs $\lambda^*$ corresponding to $\omega_R = 0$.

## 4. Element description

In the present work, $C^0$ eight-noded serendipity quadrilateral shear flexible plate element is used. The finite element represented as per the kinematics based on Equation (5), is referred as HSDT13 with cubic variation. The 13 dofs are ($u_0, v_0, w_0, \theta_x, \theta_y, w_1, \beta_x, \beta_y, \Gamma, \phi_x, \phi_y, \psi_x, \psi_y$). Four more alternate discrete models are proposed, to study the influence of the higher order terms in the displacement functions, whose displacement fields are deduced from the original element by deleting the appropriate degrees of freedom. These alternate models and the corresponding degrees of freedom are shown in Table1.

## 5. Numerical results and discussion

In this section, the flutter characteristics of sandwich plate with homogeneous core and CNT reinforced facesheets using the eight-noded shear flexible quadrilateral element is presented. The effect of various parameters such as the plate thickness and the aspect ratio, the thermal environment, the CNT volume fraction, etc. on the global response is numerically studied. Here, the sandwich plate is assumed to be simply supported and is defined as follows:

$$u_o = w_o = \theta_x = w_1 = \Gamma = \beta_x = \phi_x = \psi_x = 0 \text{ on } y=0, b$$

$$v_o = w_o = \theta_y = w_1 = \Gamma = \beta_y = \phi_y = \psi_y = 0 \text{ on } x=0, a \qquad (22)$$

where $a$ and $b$ refer to the length and width of the plate, respectively. For the present study, three different core-to-facesheet thickness $h_H/h_f$ = 8, 6, 4 and four thickness ratios $a/h$ =5, 10, 20, 50 are considered. The distribution of CNT in the facesheets is functionally graded through the thickness unless otherwise specified.

*Material properties:*

In the present investigation, Poly {(m-phenylenevinylene)-co-[(2,5-dioctoxy-p-phenylene) vinylene]}, referred as PmPV, is selected as the matrix in which the CNT's are used as reinforcements for certain cases. The material properties [8] of which are assumed to be $\rho_m = 1150$ kg/m³, $\upsilon_m = 0.34$, $\alpha_m = 45(1+0.0005\Delta T) \times 10^{-6}/K$ and $E_m = (3.51 - 0.0047 T^*)$ GPa. The temperature is defined as $T^* = T_o + \Delta T$ with $T_o = 300K$ and $\Delta T$ is the increase in temperature. Single walled CNTs are used as reinforcements and the material properties at different temperatures are given in Table 2. The CNT efficiency parameter $\eta_j$ are determined according to the effective properties of CNTRCs available by matching the Young's moduli $E_1$ and $E_2$ with the counterparts compared by the rule of mixtures [8]. The efficiency parameters are: $\eta_1 = 0.149$, $\eta_2 = 0.934$ for the case of $V_{CN}^* = 0.11$; $\eta_1 = 0.149$, $\eta_2 = 1.381$, for the case of $V_{CN}^* = 0.17$. It is assumed here as $\eta_2 = \eta_3$ and the shear moduli are assumed to be $G_{13} = G_{12} = G_{23}$. Polymethyl methacrylate (PMMA) is also considered as a candidature for matrix [28] and it is used for $V_{CN}^* = 0.28$. The Young's modulus of the matrix considered for PMMA is $E_m = (3.52 - 0.0034 T^*)$ GPa and all other properties are same as that of PMPV. The corresponding CNT efficiency factors are: $\eta_1 = 0.141$, $\eta_2 = 1.585$, $\eta_3 = 1.109$. For this case, the shear moduli are assumed to be $G_{13} = G_{12}$ and $G_{23} = 1.2 G_{12}$. Titanium alloy Ti-6Al-4V is considered for the homogeneous core in the present analysis. The properties are: $\alpha_H = 7.5788(1 + 6.638 \times 10^{-4} T^* - 3.147 \times 10^{-6} T^{*2}) \times 10^{-6} K$ Young's modulus, $E_H = 122.56(1 - 4.568 \times 10^{-4} T^*) GPa$, Poisson's ratio $\upsilon_H = 0.29$ and mass density $\rho_H = 4429 Kg/m^3$. The CNTs are either uniformly distributed or functionally graded along the thickness direction, given by

$$V_{CN}(z) = \begin{cases} V^*_{CN} & UD \\ \left(1 + \dfrac{2z}{h}\right) V^*_{CN} & FG-V \ Type \\ 2\left(\dfrac{2|z|}{h}\right) V^*_{CN} & FG-X \ Type \end{cases} \quad (25)$$

The effective material properties, viz., Young's modulus, Poisson's ratio and the mass density are estimated from Equation (1). The influence of the type of CNT volume fraction distribution in evaluating the effective properties is considered as defined in equation (25).

Table 3 presents the convergence of the critical aerodynamic pressure and the flutter frequency $\left(\Omega_i^2 = \omega^2 a^4 (\rho_H h/D_H); \lambda_{cr} = \lambda^*_{cr} a^3/D_H; D_H = E_H h^3/12(1-\nu_H^2)\right)$, for a chosen value of CNT volume fraction, with decreasing mesh size for a simply supported square sandwich plate with $a/h = 5$ and $h_H/h_h = 8$ employing both first- and higher-order (FSDT5, HSDT13) structural models. A very good convergence of results is observed with increase in the mesh discretization. For the problem considered here, a $8 \times 8$ mesh is found to be very much adequate to model the full plate, irrespective of the types of structural models.

To validate the efficacy of the present formulation, the free vibration characteristics of a single-layer carbon nanotube reinforced plate wherein the CNT is either distributed uniformly or functionally graded along the thickness as given in equation (25), is carried out and the results are shown in Table 4 for different CNT volumetric fraction and plate thickness ratio. They match very well with those of available results in literature [8]. The structural model developed here is further tested considering the flutter problem of isotropic plates and the solutions are tabulated in Tables 5. These results are again found to be in excellent agreement with those of reported in the literature [25]. Numerical experimentation is further conducted to examine the suitability of an appropriate structural theory using different structural models deduced from the present formulation as given Table 1 and the calculated results varying the thickness ratios are given in Table 6 for the selected sandwich construction and CNT volume fraction. It is noticed from Table 6 that the higher-order model HSDT11A is in close agreement with those of HSDT13. Also, it may be opined that the influence of higher-order theories is significant in particular predicting the flutter characteristics for thick sandwich plates and the results for different models approach to those of FSDT formulation for thin cases. However, further investigation here is done employing the HSDT13 and the FSDT for evaluating the behavior of CNT reinforced sandwich plates exposed to aerodynamic flow.

A detailed investigation is made to bring out the influence of the core-to-facesheet thickness ratio ($h_H/h_f$ = 8, 6, 4), the CNT volume fraction ($V^*_{CN}$ = 0.11, 17, 0.28) and the temperature ($T^*$ = 300, 500, 700)K against the sandwich plate thickness ratio on the critical aerodynamic dynamic pressure and they are depicted in Figures 2-4. It is inferred from Figure 2 that, for the given temperature, the non-dimensional critical aerodynamic pressure for a sandwich plate thickness ratio $a/h \leq 20$ predicted adopting HSDT13 model are significantly different and less compared to those of FSDT5 due to the enhanced shear flexibility associated with the HSDT13 theory. It is also revealed that increasing the homogeneous core thickness results in increase in the non-dimensional critical flutter speed. However, there is a possibility that, with the increase in CNT volume fraction, the sandwich structure with low core-to-facesheet ratio may in general predict higher flutter boundary with the increase in aspect ratio and it depends on the temperature as highlighted in Figures 3 and 4. This is mainly due to the coalescence of higher and lower modes in determining the critical flutter behavior. It may be further concluded that, with the increase in the temperature, the results evaluated by FSDT5 is significantly higher than those of HSDT13 which has greater shear flexibility and accounts for the thickness stretching mode and this trend is observed while dealing with the bending analysis of composite plate subjected to thermal loading based on higher-order model [26].

The coalescence modes that are associated with the critical flutter speed is presented in Figure 5 for both thick and thin core sandwich plates ($h_H/h_f$ = 8, 4) assuming different thickness ratio ($a/h$ = 5, 20, 50). It is clearly noticed from this Figure that the first two lower modes coalescence each other for thin plate having high core thickness whereas the lowest one coalescence with the higher mode while predicting the critical flutter speed for thick sandwich case. However, for low core thickness case, $h_H/h_f \leq 8$, the coalescence of higher modes determines the critical aerodynamic pressure, irrespective of thickness ratio of the plate considered here. This is possibly attributed to the increase in the stiffness of the facesheet of the sandwich plates.

The relative in-plane displacements and the transverse displacement through the thickness direction of the sandwich plate ($h_H/h_f$ = 8, and $a/h$ = 5, 20), for the coalescence mode of the chosen thickness ratios, are plotted in Figures 6 and 7 considering two values of flutter speed. The mode shape along the flow direction is also included in these Figures. The relative displacements ($u^*$, $v^*$, $w^*$) are plotted along the lines ($a/2$, $b/2$, $z$) where $-h/2 \leq z \leq h/2$. It is shown from these Figures that the transverse displacement $w$ is not uniform and exhibits the existence of normal stresses in the thickness direction. The variation of the transverse displacement is less at the centre of plate as the aerodynamic pressure approaches the critical value and this is attributed to the shift in the position of the maximum displacement towards the rear end of the plate. This can be seen in the flexural mode shape plot along the flow direction. It can be also viewed that the variation of in-plane displacement is not significant compared to that of transverse displacement and they are linear or nonlinear, irrespective of existence of aerodynamic flow.

For the chosen values of CNT volume fraction and temperature, the influence of the aspect ratio $a/b$ on the flutter characteristics of sandwich plates is evaluated and the results obtained here are highlighted in Table 7. It is revealed that the values of the critical aerodynamic pressure and the coalescence frequency increase with the increase in the aspect ratio. The coalescence of higher modes is in general responsible for yielding higher critical values. It can be also opined that the increase in the CNT volume fraction results in increase in the flutter speed. Lastly, the effect of aerodynamic damping is also examined assuming thick panels (a/h=5,10 and $h_H/h_f$ = 8, 4 and $T^*$=300K) and the flutter response is tabulated in Table 8. It is noticed from this Table that the introduction of damping enhances the flutter instability boundary.

Lastly, the influence of the functionally graded CNT distribution through the thickness of the facesheets over the uniform one is depicted in Table 9. It is revealed from this Table that the critical flutter speed is higher in general for the functionally graded CNT plate compared to those of the uniform case. It is further seen that the rate of increase in the critical value is more for functionally graded plate while decreasing the core thickness of sandwich plate as well as increasing the CNT volume fraction. The influence of temperature affects the performance of the sandwich plate against aerodynamic flow significantly.

## 6. Conclusions

The flutter behavior of sandwich panels with CNT reinforced facesheets are studied considering various parameters such as the sandwich type, the temperature effects, the thickness and aspect ratio, and the volume fraction of CNT. Different plate models are employed in predicting the flutter frequencies and the critical aerodynamic pressure. From a detailed investigation on the effectiveness of the chosen structural model, the following observations can be made:

(i) HSDT11A in general predicts the flutter boundary of the structure very close to the full structural model considered here, HSDT13.
(ii) The performance of the higher-order model HSDT13 for thick case is significantly different from the other lower order theories considered here and the predicted critical aerodynamic pressure is low.
(iii) Increase in the volume fraction of CNT distribution in the facesheets, in general, results in increase in the flutter boundary.
(iv) The effect of temperature based on the first-order model overestimates significantly the critical aerodynamic pressure in comparison with the higher-order one.
(v) The increase in the aspect ratio and the introduction of aerodynamic damping increases the critical flutter speed, as expected.
(vi) For thin plates, with the increase in the CNT volume fraction, the sandwich having lower core-to-facesheet thickness may predict higher critical aerodynamic pressure due to the coalescence of lower mode with the higher one.
(vii) The CNT distribution in a graded fashion through the thickness enhances the flutter boundary compared to that of uniform distribution case.
**(viii)** In-plane relative displacement has slope discontinuity at the layer interface whereas the transverse displacement varies quadratically through the thickness, as expected**.**
(ix) The occurrence of type of flexural/extensional modes in the thickness direction depends on the location of the structures and the flexural mode along the length of the plate that corresponds to coalescence changes its shape as the airflow is introduced.

**Acknowledgements**

S Natarajan would like to acknowledge the financial support of the School of Civil and Environmental Engineering, The University of New South Wales, Sydney, for his research fellowship for the period September 2012 onwards. The authors would also to acknowledge the contributions of MP Mathyarasy, postgraduate student, Hindustan University, in carrying out some of the parametric studies.

**Table 1:** Alternate eight-noded finite element structural models

| Finite element model | Degrees of freedom per node |
|---|---|
| HSDT13 | $u_o, v_o, w_o, \theta_x, \theta_y, w_1, \beta_x, \beta_y, \Gamma, \phi_x, \phi_y, \psi_x, \psi_y$ |
| HSDT11A | $u_o, v_o, w_o, \theta_x, \theta_y, \beta_x, \beta_y, \phi_x, \phi_y, \psi_x, \psi_y$ |
| HSDT11B | $u_0, v_0, w_0, \theta_x, \theta_y, w_1, \beta_x, \beta_y, \Gamma, \phi_x, \phi_y$ |
| HSDT9 | $u_0, v_0, w_0, \theta_x, \theta_y, \beta_x, \beta_y, \phi_x, \phi_y$ |
| FSDT5 | $u_0, v_0, w_0, \theta_x, \theta_y$ |

**Table 2:** Temperature dependent material properties for (10, 10) SWCNT [28]

| Temperature K | $E_{11}^{CN}$ (Tpa) | $E_{22}^{CN}$ (Tpa) | $G_{12}^{CN}$ (Tpa) | $\alpha_{11}^{CN}$ ($\times 10^{-6}$ K) | $\alpha_{22}^{CN}$ ($\times 10^{-6}$ K) |
|---|---|---|---|---|---|
| 300 | 5.6466 | 7.0800 | 1.9445 | 3.4584 | 5.1682 |
| 500 | 5.5308 | 6.9348 | 1.9643 | 4.5361 | 5.0189 |
| 700 | 5.4744 | 6.8641 | 1.9644 | 4.6677 | 4.8943 |

**Table 3:** Convergence of critical aerodynamic pressure, $\lambda_{cr}$ with mesh size for a square sandwich plate with $a/h$=5. The volume fraction of the CNT is $V_{CN}^*$=0.11, and Temperature, $T^*$ = 300.
$\left( \Omega_i^2 = \omega^2 a^4 \left( \rho_H h / D_H \right); \lambda_{cr} = \lambda_{cr}^* a^3 / D_H; D_H = E_H h^3 / 12(1 - \nu_H^2) \right)$

| Mesh | $h_H/h_f$ | Plate Theories | In-vacuo | | | | Coalescence | |
|---|---|---|---|---|---|---|---|---|
| | | | $\Omega_1^2$ | $\Omega_2^2$ | $\Omega_3^2$ | $\Omega_4^2$ | $\Omega_{cr}^2$ | $\lambda_{cr}$ |
| 2×2 | 8 | HSDT13 | 253.46 | 906.74 | 1004.20 | 1004.20 | 691.39 | 141.99 |
| | | FSDT5 | 256.97 | 887.55 | 1004.30 | 1004.30 | 696.96 | 145.89 |
| 4×4 | 8 | HSDT13 | 235.75 | 997.19 | 997.19 | 1092.60 | 1395.55 | 305.86 |
| | | FSDT5 | 238.87 | 997.33 | 997.33 | 1087.90 | 1445.40 | 329.49 |
| 6×6 | 8 | HSDT13 | 234.81 | 996.78 | 996.78 | 1077.20 | 1297.91 | 290.82 |
| | | FSDT5 | 237.90 | 996.92 | 996.92 | 1072.90 | 1305.52 | 306.83 |
| 8×8 | 8 | HSDT13 | 234.64 | 996.71 | 996.71 | 1073.80 | 1291.56 | 289.84 |
| | | FSDT5 | 237.73 | 996.85 | 996.85 | 1069.60 | 1295.16 | 305.08 |
| 16×16 | 8 | HSDT13 | 234.64 | 996.71 | 996.71 | 1073.80 | 1291.56 | 289.84 |
| | | FSDT5 | 237.73 | 996.85 | 996.85 | 1069.60 | 1295.16 | 305.08 |

**Table 4:** Comparison of fundamental natural vibration frequency ($\bar{\omega} = \omega(a^2/h)\sqrt{\rho_m/E_m}$) of simply supported CNTRC square plate with different CNT volume fraction, $V_{CN}^*$, thickness ratio $a/h$, and distribution of CNT (Uniform-UD, Functionally Graded –V & X)

| $V_{CN}^*$ | $a/h$ | Non-dimensional natural frequency, $\bar{\omega}$ | | | | | |
|---|---|---|---|---|---|---|---|
| | | UD | | FG-V | | FG-X | |
| | | Present | Ref. [8] | Present | Ref. [8] | Present | Ref. [8] |
| 0.11 | 10 | 13.696 | 13.532 | 12.564 | 12.452 | 14.817 | 14.616 |
| | 20 | 17.411 | 17.355 | 15.136 | 15.110 | 20.030 | 19.939 |
| | 50 | 19.187 | 19.223 | 16.222 | 16.252 | 22.840 | 22.984 |
| 0.14 | 10 | 14.510 | 14.306 | 13.402 | 13.256 | 15.603 | 15.368 |
| | 20 | 19.027 | 18.921 | 16.573 | 16.510 | 21.791 | 21.642 |
| | 50 | 21.357 | 21.354 | 17.996 | 17.995 | 25.560 | 25.555 |
| 0.17 | 10 | 17.000 | 16.815 | 15.585 | 15.461 | 18.498 | 18.278 |
| | 20 | 21.513 | 21.456 | 18.662 | 18.638 | 24.849 | 24.764 |
| | 50 | 23.645 | 23.697 | 19.945 | 19.982 | 28.341 | 28.413 |

**Table 5:** Critical aerodynamic pressure and coalescence frequency for an isotropic square plate $\left(\bar{\Omega}_{cr}^2 = \omega^2 a^4 (\rho\, h/D);\ \bar{\lambda}_{cr} = \lambda_{cr}^* a^3/D\ ;\ D = E\, h^3/12(1-\nu^2)\right)$

| | Simply supported plate | | Clamped plate | |
|---|---|---|---|---|
| | $\bar{\lambda}_{cr}$ | $\bar{\Omega}_{cr}^2$ | $\bar{\lambda}_{cr}$ | $\bar{\Omega}_{cr}^2$ |
| Present | 511.11 | 1840.29 | 852.34 | 4274.32 |
| Ref. [25] | 512.33 | 1846.55 | 852.73 | 4294.07 |

**Table 6:** Flutter boundary of CNT reinforced square sandwich plates based on different finite element models. The volume fraction of the CNT, $V_{CN}^*$=0.11 and CNT is functionally graded, and Temperature, $T^*$=300K.

| $a/h$ | $h_H/h_f$ | Plate theories | In vacuo | | | | In coalescence | |
|---|---|---|---|---|---|---|---|---|
| | | | $\Omega_1^2$ | $\Omega_2^2$ | $\Omega_3^2$ | $\Omega_4^2$ | $\Omega_{cr}^2$ | $\lambda_{cr}$ |
| 5 | 8 | HSDT13 | 234.80 | 996.78 | 996.78 | 1077.2 | 1297.91 | 290.82 |
| | | HSDT11A | 233.30 | 996.78 | 996.78 | 1060.2 | 1236.85 | 284.96 |
| | | HSDT11B | 240.72 | 996.78 | 996.78 | 1082.1 | 1370.89 | 316.99 |
| | | HSDT9 | 239.25 | 996.78 | 996.78 | 1065.3 | 1312.76 | 312.30 |
| | | FSDT | 237.90 | 996.92 | 996.92 | 1072.9 | 1305.52 | 306.84 |
| 10 | 8 | HSDT13 | 272.28 | 1394.54 | 1892.90 | 3811.66 | 1534.11 | 370.11 |
| | | HSDT11A | 271.78 | 1387.17 | 1886.92 | 3786.83 | 1517.92 | 369.14 |
| | | HSDT11B | 274.48 | 1397.06 | 1993.66 | 3909.51 | 1556.95 | 381.44 |
| | | HSDT9 | 273.98 | 1389.70 | 1988.10 | 3885.30 | 1541.93 | 380.85 |
| | | FSDT | 273.53 | 1392.80 | 1965.90 | 3862.90 | 1537.20 | 378.32 |
| 20 | 8 | HSDT13 | 284.14 | 1513.20 | 2169.50 | 4433.30 | 1624.34 | 402.15 |
| | | HSDT11A | 284.00 | 1510.90 | 2167.60 | 4424.80 | 1620.57 | 401.95 |
| | | HSDT11B | 284.75 | 1514.00 | 2204.90 | 4470.20 | 1628.14 | 405.08 |
| | | HSDT9 | 284.61 | 1511.70 | 2203.10 | 4461.80 | 1624.33 | 404.88 |
| | | FSDT | 284.48 | 1512.60 | 2196.00 | 4454.00 | 1623.56 | 404.29 |
| 50 | 8 | HSDT13 | 287.69 | 1551.40 | 2268.20 | 4657.50 | 1643.81 | 410.55 |
| | | HSDT11A | 287.66 | 1550.70 | 2267.80 | 4655.60 | 1643.45 | 410.55 |
| | | HSDT11B | 287.80 | 1551.50 | 2274.60 | 4664.20 | 1642.95 | 410.74 |
| | | HSDT9 | 287.76 | 1550.90 | 2274.10 | 4662.40 | 1642.53 | 410.74 |
| | | FSDT | 287.74 | 1551.02 | 2272.90 | 4661.00 | 1642.88 | 410.74 |

**Table 7:** Flutter boundary of rectangular CNT reinforced sandwich plates ($a/b$=1, 2 and 5). The volume fraction of the CNT, $V^*_{CN}$=0.17; CNT is functionally graded and the temperature, $T^*$=300K

| $a/h$ | $a/b$ | Plate theory | In vacuo | | | | Coalescence | |
|---|---|---|---|---|---|---|---|---|
| | | | $\Omega_1^2$ | $\Omega_2^2$ | $\Omega_3^2$ | $\Omega_4^2$ | $\Omega_{cr}^2$ | $\lambda_{cr}$ |
| 5 | 1 | HSDT13 | 252.66 | 999.00 | 999.00 | 1098.50 | 1489.52 | 332.23 |
| | | FSDT5 | 256.68 | 999.13 | 999.13 | 1093.00 | 1509.34 | 353.32 |
| | 2 | HSDT13 | 999.00 | 1094.20 | 2643.10 | 3994.50 | 3261.81 | 526.56 |
| | | FSDT5 | 999.13 | 1088.90 | 2698.80 | 3996.50 | 3014.10 | 514.45 |
| | 5 | HSDT13 | 999.00 | 4000.40 | 9048.50 | 13536.0 | 20779.45 | 1285.74 |
| | | FSDT5 | 999.13 | 4002.40 | 9058.90 | 13358.0 | 19792.06 | 1248.05 |
| 10 | 1 | HSDT13 | 297.08 | 1428.50 | 2183.30 | 3996.40 | 1784.80 | 434.57 |
| | | FSDT5 | 298.70 | 1426.10 | 2279.10 | 3996.50 | 1792.08 | 445.70 |
| | 2 | HSDT13 | 1421.40 | 3996.40 | 4057.10 | 10821.0 | 3860.48 | 680.27 |
| | | FSDT5 | 1419.10 | 3996.50 | 4123.70 | 11421.0 | 3758.02 | 677.15 |
| | 5 | HSDT13 | 3996.40 | 16008.0 | 26387.0 | 32136.0 | 36854.20 | 2275.85 |
| | | FSDT5 | 3996.50 | 16010.0 | 26200.0 | 31887.0 | 35773.78 | 2240.82 |

**Table 8:** Aerodynamic damping on critical aerodynamic pressure of CNT reinforced square sandwich plate. The damping coefficient, $g_T$= 0.1 and CNT is functionally graded and temperature, $T^*$=300K

| $V^*_{CN}$ | $a/h$ | $h_H/h_f$ | Plate theory | In vacuo | | | | No Damping | | With Damping | |
|---|---|---|---|---|---|---|---|---|---|---|---|
| | | | | $\Omega_1^2$ | $\Omega_2^2$ | $\Omega_3^2$ | $\Omega_4^2$ | $\Omega_{cr}^2$ | $\lambda_{cr}$ | $\Omega_1^2$ | $\lambda_{cr}$ |
| 0.17 | 5 | 8 | HSDT13 | 252.66 | 999.00 | 999.00 | 1098.50 | 1489.52 | 332.23 | 1572.30 | 353.20 |
| | | | FSDT5 | 256.68 | 999.13 | 999.13 | 1093.00 | 1509.34 | 353.32 | 1577.34 | 372.97 |
| | | 4 | HSDT13 | 211.32 | 838.57 | 946.90 | 946.90 | 1287.02 | 252.15 | 1354.88 | 266.72 |
| | | | FSDT5 | 234.87 | 846.67 | 947.39 | 947.39 | 1647.81 | 342.97 | 1721.26 | 360.25 |
| | 10 | 8 | HSDT13 | 297.08 | 1428.50 | 2183.30 | 3996.40 | 1784.80 | 434.57 | 1832.16 | 452.89 |
| | | | FSDT5 | 298.70 | 1426.10 | 2279.10 | 3996.50 | 1792.08 | 445.70 | 1832.88 | 463.24 |
| | | 4 | HSDT13 | 264.02 | 1056.50 | 2073.10 | 3424.40 | 1808.93 | 388.48 | 1862.17 | 404.14 |
| | | | FSDT5 | 275.19 | 1062.70 | 2503.20 | 3674.30 | 1974.29 | 446.09 | 2014.94 | 461.21 |
| 0.28 | 5 | 8 | HSDT13 | 275.17 | 998.72 | 998.72 | 1126.60 | 1711.76 | 376.56 | 1823.32 | 402.62 |
| | | | FSDT5 | 281.20 | 998.85 | 998.85 | 1107.40 | 1804.08 | 414.06 | 1904.68 | 439.73 |
| | | 4 | HSDT13 | 237.66 | 879.72 | 946.53 | 946.53 | 1477.14 | 284.96 | 1566.77 | 302.97 |
| | | | FSDT5 | 275.18 | 888.92 | 947.01 | 947.01 | 2120.71 | 423.44 | 2249.39 | 448.87 |
| | 10 | 8 | HSDT13 | 333.10 | 1477.50 | 2552.70 | 3995.30 | 2145.72 | 523.83 | 2209.05 | 546.76 |
| | | | FSDT5 | 335.83 | 1468.70 | 2714.60 | 3995.40 | 2176.73 | 544.92 | 2230.94 | 566.99 |
| | | 4 | HSDT13 | 317.21 | 1134.40 | 2487.60 | 3787.50 | 2291.11 | 486.52 | 2372.17 | 508.05 |
| | | | FSDT5 | 337.91 | 1144.70 | 3176.70 | 3788.00 | 2615.56 | 588.87 | 2678.93 | 610.16 |

**Table 9:** Flutter behavior of sandwich square plates with different distribution of CNT through the thickness of facesheet (uniform –UD and Functionally Graded-FG).

| $a/h$ | $V_{CN}^*$ | $h_H/h_f$ | $T^*$ | CNT | In-vacuo | | | | Coalescence | |
|---|---|---|---|---|---|---|---|---|---|---|
| | | | | | $\Omega_1^2$ | $\Omega_2^2$ | $\Omega_3^2$ | $\Omega_4^2$ | $\Omega_{cr}^2$ | $\lambda_{cr}$ |
| 5 | 0.17 | 8 | 300 | UD | 252.14 | 998.89 | 998.89 | 1097.2 | 1497.35 | 337.70 |
|   |      |   |     | FG | 252.66 | 999.00 | 999.00 | 1098.5 | 1489.52 | 332.23 |
|   |      |   | 700 | UD | 186.69 | 781.01 | 781.01 | 841.45 | 1008.24 | 224.80 |
|   |      |   |     | FG | 180.81 | 781.03 | 781.03 | 837.43 | 929.85  | 205.86 |
|   |      | 4 | 300 | UD | 212.55 | 837.92 | 946.69 | 946.69 | 1348.93 | 268.94 |
|   |      |   |     | FG | 211.32 | 838.57 | 946.90 | 946.90 | 1287.02 | 252.15 |
|   |      |   | 700 | UD | 139.11 | 611.29 | 735.08 | 735.08 | 771.18  | 155.86 |
|   |      |   |     | FG | 130.28 | 604.56 | 735.09 | 735.09 | 672.98  | 134.57 |
|   | 0.28 | 8 | 300 | UD | 275.11 | 998.24 | 998.24 | 1123.5 | 1738.67 | 387.70 |
|   |      |   |     | FG | 275.17 | 998.72 | 998.72 | 1126.6 | 1711.76 | 376.56 |
|   |      |   | 700 | UD | 218.72 | 783.02 | 783.02 | 876.08 | 1352.22 | 298.83 |
|   |      |   |     | FG | 215.10 | 783.24 | 783.24 | 874.88 | 1270.17 | 277.34 |
|   |      | 4 | 300 | UD | 241.00 | 876.78 | 945.61 | 945.62 | 1579.72 | 310.74 |
|   |      |   |     | FG | 237.66 | 879.72 | 946.53 | 946.53 | 1477.14 | 284.96 |
|   |      |   | 700 | UD | 183.49 | 667.18 | 739.01 | 739.02 | 1126.63 | 222.65 |
|   |      |   |     | FG | 174.56 | 662.75 | 739.43 | 739.43 | 987.55  | 192.19 |
| 20 | 0.17 | 8 | 300 | UD | 306.61 | 1546.80 | 2498.90 | 4779.40 | 1851.82 | 465.04 |
|   |      |   |     | FG | 311.39 | 1552.70 | 2554.30 | 4836.00 | 1902.81 | 477.93 |
|   |      |   | 700 | UD | 245.83 | 1198.10 | 1953.70 | 3705.10 | 1511.28 | 375.19 |
|   |      |   |     | FG | 247.66 | 1200.00 | 1901.80 | 3654.00 | 1510.21 | 370.90 |
|   |      | 4 | 300 | UD | 270.62 | 1120.20 | 2585.60 | 4144.60 | 1921.72 | 437.50 |
|   |      |   |     | FG | 283.50 | 1135.30 | 2707.90 | 4168.20 | 2053.09 | 465.62 |
|   |      |   | 700 | UD | 211.19 | 840.80  | 1746.10 | 2911.60 | 1415.05 | 305.96 |
|   |      |   |     | FG | 213.32 | 843.11  | 1619.80 | 2785.80 | 1375.90 | 294.33 |
|   | 0.28 | 8 | 300 | UD | 345.21 | 1599.40 | 5336.60 | 2254.95 | 2254.95 | 574.61 |
|   |      |   |     | FG | 352.82 | 1611.60 | 5420.10 | 2331.90 | 2331.90 | 593.55 |
|   |      |   | 700 | UD | 286.95 | 1256.80 | 2601.60 | 4374.30 | 1953.28 | 498.63 |
|   |      |   |     | FG | 293.21 | 1265.10 | 2635.50 | 4410.90 | 2014.54 | 511.91 |
|   |      | 4 | 300 | UD | 331.30 | 1204.60 | 3374.80 | 4324.00 | 2544.67 | 584.37 |
|   |      |   |     | FG | 351.09 | 1233.60 | 3529.80 | 4387.80 | 2736.50 | 623.63 |
|   |      |   | 700 | UD | 281.64 | 940.77  | 2829.90 | 3295.30 | 2220.24 | 504.69 |
|   |      |   |     | FG | 296.67 | 959.78  | 2854.20 | 3329.80 | 2349.48 | 525.20 |

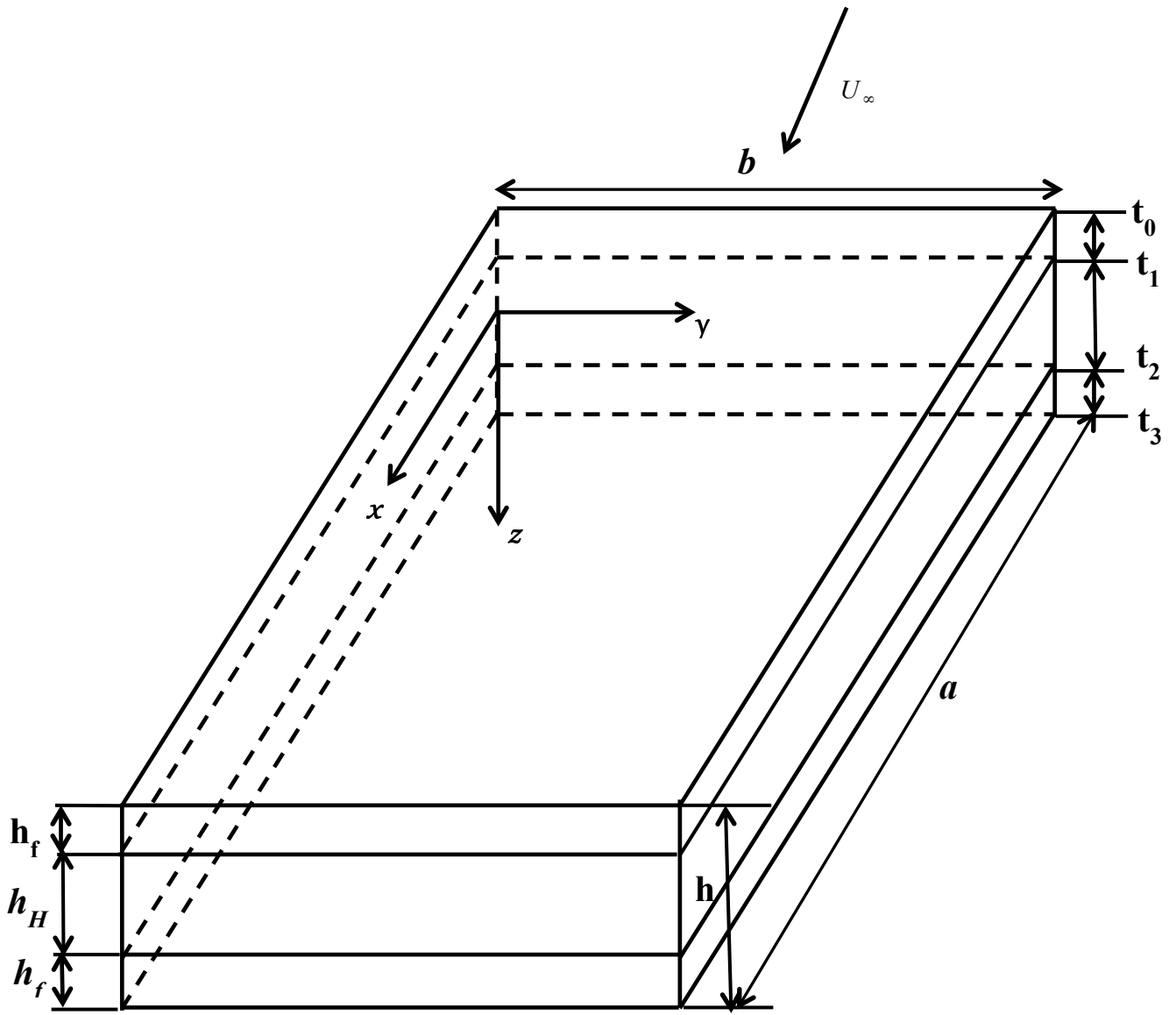

**Figure1**. Coordinate system of a rectangular sandwich plate with *x* and *y* along the in-plane directions and *z* along the plane cross section.

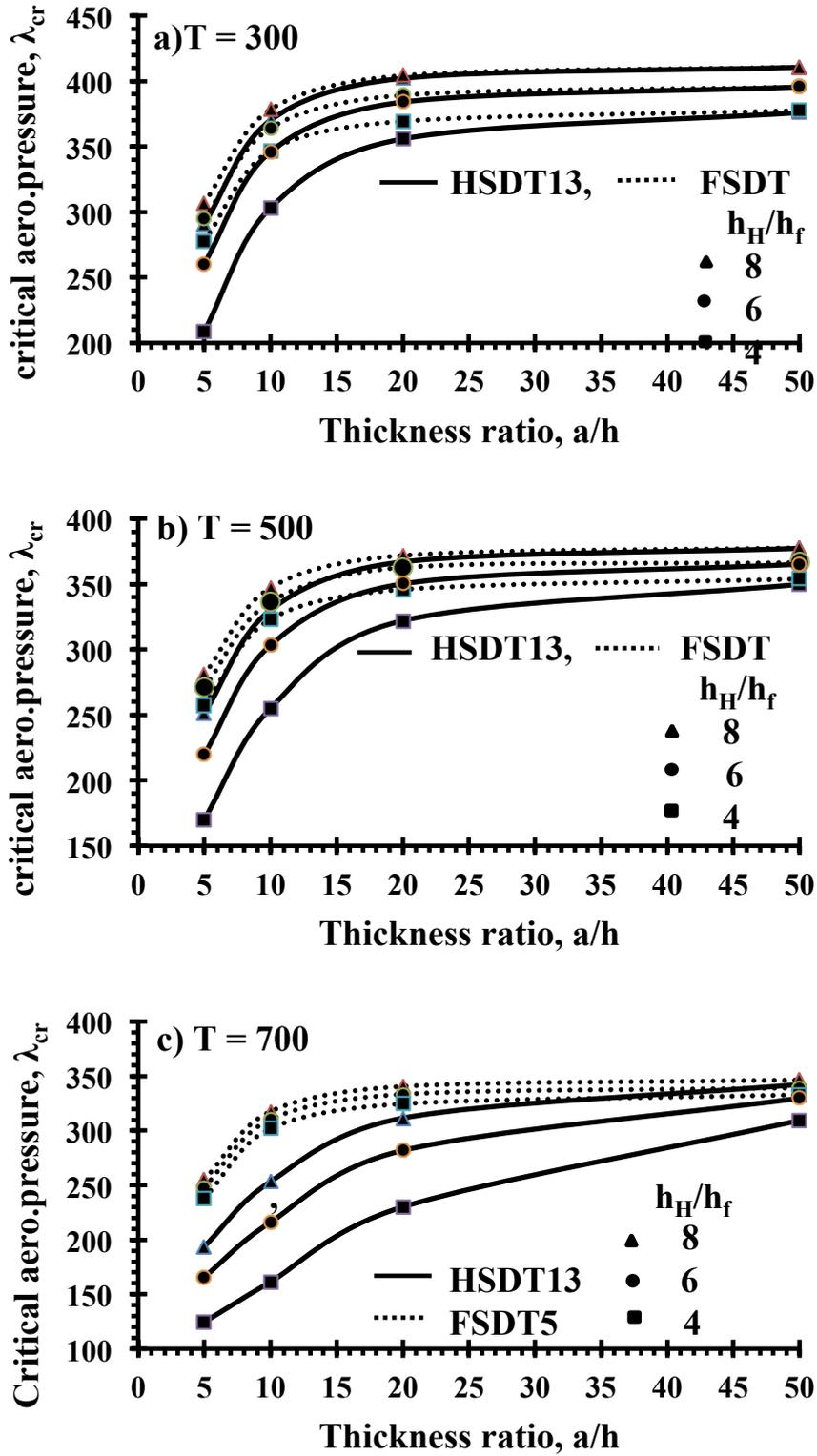

**Figure 2.** Variation of the critical aerodynamic pressure with aspect ratio for a square sandwich plate with different core-to-facesheet and temperature ($V^*_{CN}$ =0.11).

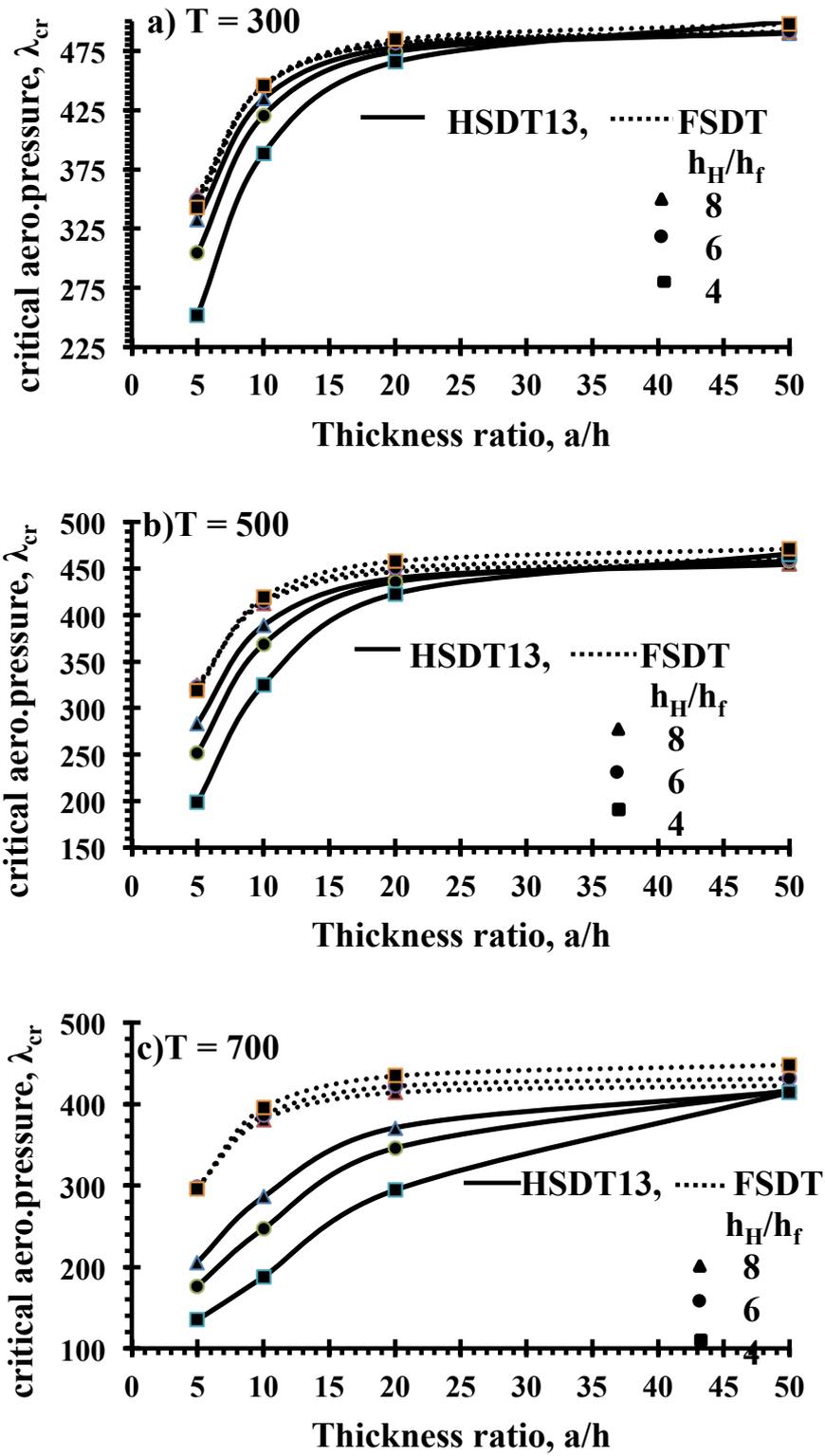

**Figure 3.** The variation of critical aerodynamic pressure with aspect ratio for a square sandwich plate with different core-to- facesheet and temperature ($V^*_{CN}$ =0.17).

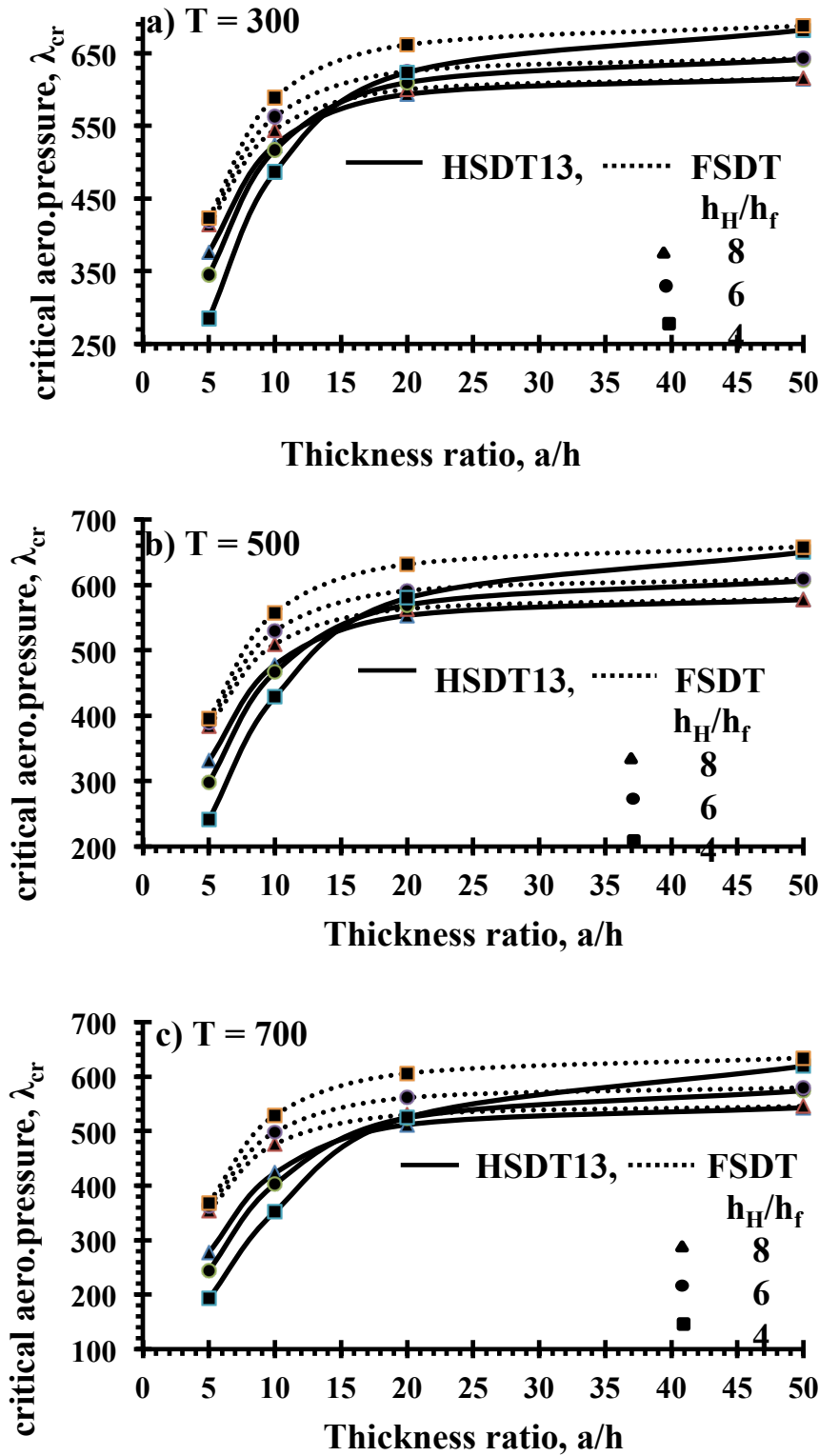

**Figure 4.** The variation of critical aerodynamic pressure with aspect ratio for a square sandwich plate with different core-to- facesheet and temperature ($V^*_{CN}$ =0.28).

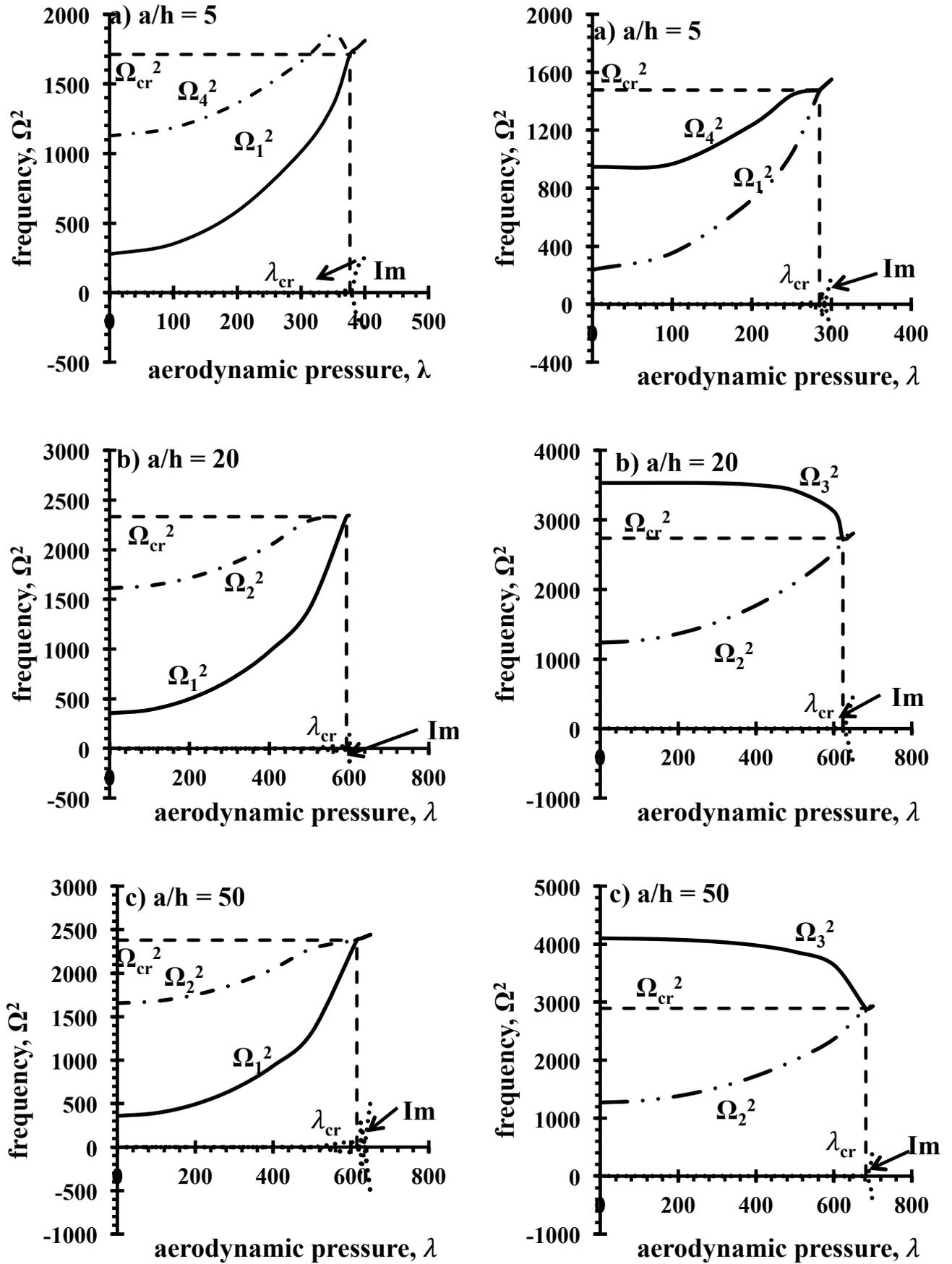

**Figure 5.** The coalesence modes corresponding to critical flutter speed for different thickness ratio of a square sandwich plate($V^*_{CN}$ =0.28, $T^*$=300): (a) Left side, $h_H/h_f$=8, (b) Right side, $h_H/h_f$ =4

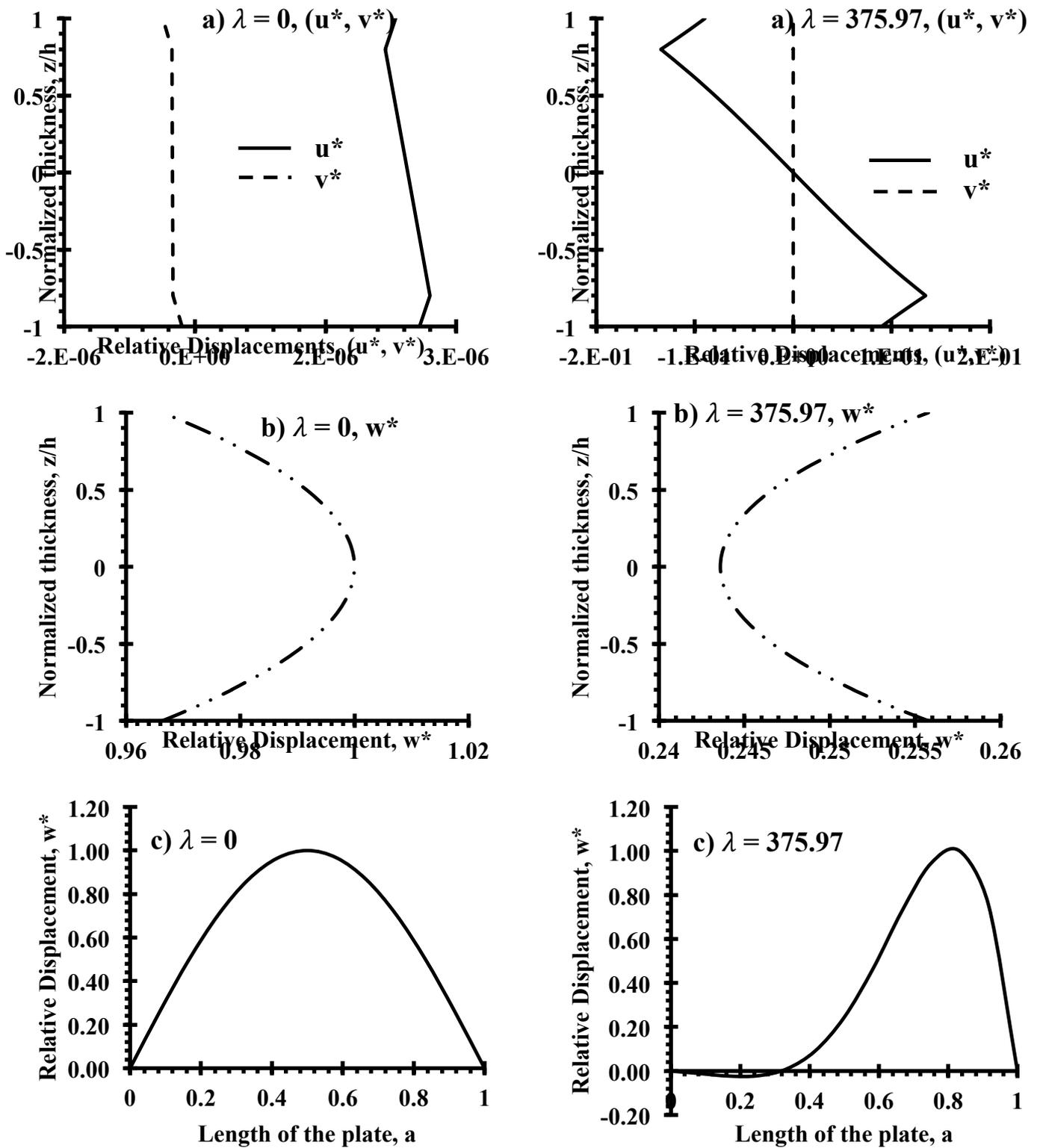

**Figure 6**. The relative displacements of (u, v, w) through thickness, and transverse displacement along the air flow of a square plate (a/h=5, $V^*_{CN}$ =0.28, $T^*$=300): (a) Left side $\lambda$=0, (b) Right side $\lambda$=375.97

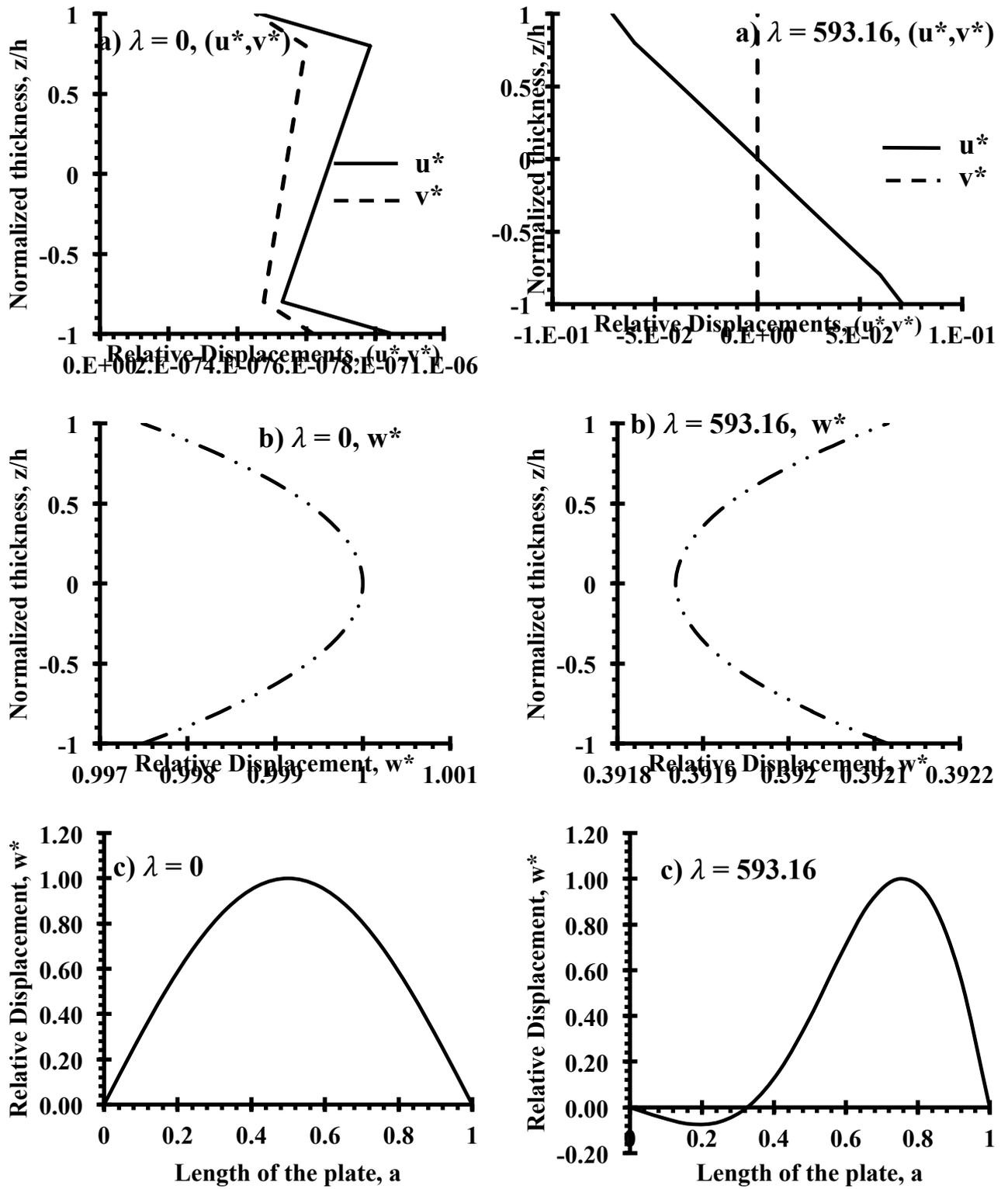

**Figure 7.** The relative displacements of (u, v, w) through thickness, and transverse displacement along the air flow of a square plate (a/h=20, $V^*_{CN}$ =0.28, $T^*$=300): (a) Left side $\lambda$=0, (b) Right side $\lambda$=593.16